# Spin Diffusion and Relaxation in a Nonuniform Magnetic Field.

G.P. Berman, B. M. Chernobrod, V.N. Gorshkov,

Theoretical Division, Los Alamos National Laboratory, Los Alamos, New Mexico 87545

V.I. Tsifrinovich

IDS Department, Polytechnic University, Brooklyn, NY 11201

## Abstract

We consider a quasiclassical model that allows us to simulate the process of spin diffusion and relaxation in the presence of a highly nonuniform magnetic field. The energy of the slow relaxing spins flows to the fast relaxing spins due to the dipole-dipole interaction between the spins. The magnetic field gradient suppresses spin diffusion and increases the overall relaxation time in the system. The results of our numerical simulations are in a good agreement with the available experimental data.

## 1. Introduction

Application of any quantum spin device crucially depends on the relaxation rate in the spin system. Spin diffusion is recognized as one of the most important factors in a spin relaxation process (see, for example, [1-6]).

The idea of spin diffusion originated from Bloembergen who explained nuclear spin-lattice relaxation in insulating crystals [1]. He demonstrated that the transport of magnetization from fast relaxing spins (FRS) to slow relaxing spins (SRS) can be described as a diffusion process. Due to the spin diffusion, a small amount of FRS (e.g. located near the impurities) can greatly accelerate the spin-lattice relaxation in the whole spin system. A number of theoretical approaches have been used to describe the spin diffusion. Redfield [2] has found the connection between the diffusion constant and the moments of the NMR line. Yang and Honig [3] applied the spin diffusion theory to electron paramagnetic system. Rhim, Pines, and Waugh [4] showed that the spin diffusion can be reversed using *rf*-pulses which change the sign of the dipole-dipole interaction. Vugmeister [5] has developed the general theory of the spin diffusion in a delute electron paramagnetic system in terms of the Zeeman spin temperatures for FRS and SRS. Tang and Waugh [6] simulated the nuclear spin diffusion in insulated crystals using classical equations of motion for the nuclear spins.

Recently Budakian, Mamin, and Rugar [7] demonstrated effective manipulation with the electron spin relaxation of $E'$- centers in silica using the high gradient of the magnetic field produced by a ferromagnetic particle in the magnetic resonance force microscopy. This effect opens the way for effective suppression of the spin diffusion and relaxation, which could be extremely important for the spin device applications. Inspired by the experiments [7] we have suggested a computational model, which could be applied to the simulation of the spin relaxation in a highly nonuniform magnetic field.

According to the general theory of the spin diffusion in the electron spin resonance [5] the relaxation process depends on the relation between two parameters: the spin-lattice relaxation time for FRS $T_{FL}$ and the cross relaxation time $T_{FS}$, which is the characteristic time of the energy transfer from FRS to SRS.

If $T_{FS} \ll T_{FL}$ then FRS and SRS quickly come to the state of thermal equilibrium, and the bottleneck of the relaxation process in the energy transfer from FRS to the lattice. In this case the overall relaxation rate $T_1^{-1}$ is given by the parameter $T_0^{-1}$: $T_1^{-1} \approx T_0^{-1} = (n_F/n_S)T_{FL}^{-1}$, where $n_F$ and $n_S$ are the concentrations of FRS and SRS. (The spin-lattice relaxation of SRS with the characteristic time $T_{SL}$ is ignored in [5].) In the opposite case $T_{FS} \gg T_{FL}$, FRS quickly come to the thermal equilibrium with the lattice, and the bottleneck of the relaxation process is the energy transfer from SRS to FRS with the characteristic cross-relaxation time $T_{SF}$: $T_{SF} = (n_S/n_F)T_{FS}$. In this case the relaxation rate $T_1^{-1}$ is, approximately, equal to $T_{SF}^{-1}$. For spins $1/2$ the value of $T_{FS}^{-1}$ is given by

$$T_{FS}^{-1} = q\left(\frac{\mu_0}{4\pi}n_S\hbar\gamma^2\right)^2,$$

where $q \approx 9.6/\delta$ for the inhomogeneously broadened resonance line with the width $\delta$, and $q \approx 4/\Lambda$ for the homogeneously broadened resonance line with the width $\Lambda$ of the cross-relaxation form function. Taking parameters from [7] $n_S = 2\times10^{24}m^{-3}$, $\delta/\gamma = 10^{-4}T$, $\gamma/(2\pi) = 2.8 \times 10^{10} Hz/T$, we estimate the value of $T_{FS}^{-1}$: $T_{FS}^{-1} = 2.3 \times 10^5 s^{-1}$. It follows from the experiments [7] that the spin-lattice relaxation rate for SRS $T_{SL}^{-1}$, which can be revealed under the high magnetic field gradient is about $0.04 s^{-1}$. It is reasonable to assume that the ratio $T_{SL}/T_{FL}$ does not exceed $10^5$, i.e. $T_{FL}^{-1} \leq 4000 s^{-1}$. In this case $T_{FS} \ll T_{FL}$, and the bottleneck of the relaxation process is the energy transfer from FRS to the lattice. For this situation, we simulated the electron spin dynamics in the presence of the nonuniform magnetic field. We consider the motion of electron spins coupled via the dipole-dipole interaction using the classical equations of motion. In the second section we describe our model and computational algorithm, and in the third section present the results of our numerical simulations and comparison with the experimental data.

## 2. Description of the model and the computational algorithm.

We consider a quasiclassical electron "spin" with magnetic moment $\vec{\mu}_p$ and negative electron gyromagnetic ratio $(-\gamma)$. The magnitude of the magnetic moment, which is equal to the Bohr's magnetron ($\mu_p = \mu_B$), conserves in the process of the spin relaxation. The motion of a magnetic moment $\vec{\mu}_p$ satisfies to the quasiclassical equation of motion with the relaxation term $\vec{R}_p$:

$$\dot{\vec{\mu}}_p = -\gamma[\vec{\mu}_p \times \vec{B}_p] + \vec{R}_p,$$
$$\vec{R}_p = \frac{\xi_p}{\mu_B}[\vec{\mu}_p \times \dot{\vec{\mu}}_p]. \qquad (1)$$

Here $\vec{B}_p$ is the magnetic field on spin "$p$", which includes the uniform external magnetic field $\vec{B}_{ext}$, the nonuniform external magnetic field produced, for example, by a ferromagnetic particle $\vec{B}_{fp}$, and the dipole field produced by other spins of a sample $\vec{B}_{dp}$; $\xi_p$ is the relaxation parameter. We assume that $\vec{B}_{ext}$ points in the positive $z$-direction.

The dipole field $\vec{B}_{dp}$ is given by

$$\vec{B}_{dp} = \frac{\mu_0}{4\pi}\sum_{k \neq p}\frac{3(\vec{\mu}_k \cdot \vec{n}_{kp}) - \vec{\mu}_k}{r_{kp}^3}, \qquad (2)$$

where $\vec{n}_{kp}$ is the unit vector, which points from the spin $k$ to the spin $p$, $r_{kp}$ is the distance between the two spins. Below we use two conditions, which were satisfied in experiments [7]:

$$B_{dp}, \ \left|\vec{B}_{fp} - \langle \vec{B}_f \rangle\right| \ll B_{ext} + \langle B_{fz} \rangle$$

where $\langle ... \rangle$ means the average over the spin system.

Note that according to the Maxwell's equation $div\vec{B}_f = 0$ there must be the nonuniform transversal component of the magnetic field $\vec{B}_f$. However because of $\left|\vec{B}_{fp} - \langle \vec{B}_f \rangle\right| \ll B_{ext} + \langle B_{fz} \rangle$ the transversal component of $\vec{B}_f$ in the first approximation does not influence the Larmor frequency.

The average precession frequency $\omega_0$ in the spin system can be found approximately as

$$\omega_0 = \gamma B_0, \quad B_0 = B_{ext} + \langle B_{fz} \rangle, \tag{3}$$

We transfer to the system of coordinates, which rotates with the frequency $\omega_0$. Ignoring the fast oscillating terms and assuming $\xi_p \ll 1$ we obtain the following equations of motion:

$$\frac{d\vec{m}_p}{d\tau} = [\vec{\Omega}_p \times \vec{m}_p],$$

$$\vec{\Omega}_p = \left( -\frac{\chi}{2}\sum_{k \neq p} A_{kp} m_{kx} + \xi'_p m_{py}, \ -\frac{\chi}{2}\sum_{k \neq p} A_{kp} m_{ky} - \xi'_p m_{px}, \ \delta_p + \chi \sum_{k \neq p} A_{kp} m_{kz} \right). \tag{4}$$

Here we use the following dimensionless notations:

$$\vec{m}_p = \vec{\mu}_p / \mu_B, \quad \dot{\vec{m}}_p = d\vec{m}_p / d\tau, \quad \tau = \xi_0 \omega_0 t, \quad \xi'_p = \xi_p / \xi_0,$$

$$\delta_p = \frac{B_{fpz} - \langle B_{fz} \rangle}{\xi_0 B_0}, \quad \chi = \frac{\mu_0}{4\pi} \frac{\mu_B}{\xi_0 a^3 B_0}, \quad A_{kp} = \frac{3\cos^2 \theta_{kp} - 1}{(r_{kp}/a)^3}, \tag{5}$$

where $a$ is the average distance between the spins, $\theta_{kp}$ is the polar angle of the unit vector $\vec{n}_{kp}$, $\xi_0 \omega_0 = 1 s^{-1}$. If we put origin inside the spin system at the point where $\delta_p = 0$ then the value of $\delta_p$ can be approximated as

$$\delta_p = \frac{aG}{\xi_0 B_0}(z_p/a), \quad G = \left.\frac{\partial B_{fz}}{\partial z}\right|_{z=0}. \tag{6}$$

The parameter $\chi$ is the characteristic constant of the dipole-dipole interaction in our system, while the ratio $G' = aG/(\xi_0 B_0)$ describes the characteristic Larmor frequency difference caused by the magnetic field gradient.

The approximations leading to (4) from (1) maintain, as it should be, the conservation of the magnetic moment:

$$m_{px}^2 + m_{py}^2 + m_{pz}^2 = 1. \tag{7}$$

In the absence of relaxation ($\xi'_p = 0$) the $z$-component of the total magnetic moment also conserves: $\sum_p \dot{m}_{pz} = 0$.

In experiments [7] the values of parameters are the following: $B_0 = 0.106\ T$, $\omega_0/2\pi = 2.96\ GHz$, $a \approx 8\ nm$, $G \approx 10\ T/m \div 3.6 \times 10^4\ T/m$, $T_1 \approx 6.25\ s \div 25\ s$. The value of $\chi$ is $\chi = 3.2 \times 10^5$. When $G$ changes from $10\ T/m$ to $3.6 \times 10^4\ T/m$ the dimensionless parameter $G'$ raises from $0.044\chi$ to $156\chi$ (for $a = 8\ nm$). Thus, the Larmor frequency difference for the neighboring spins becomes greater than the dipole-dipole interaction constant.

The computational scheme for eq. (4) must satisfy to the conditions of conservation $m_p$ and (in the absence of relaxation) $\sum_p m_{pz}$. Below we describe the suggested computational algorithm. Equations (4) are approximated by the difference equations

$$\frac{\vec{m}_p^{j+1} - \vec{m}_p^j}{\Delta \tau} = \left[\vec{\Omega}_p\left(\vec{m}_1^{(av)}, \vec{m}_2^{(av)}, \ldots \vec{m}_N^{(av)}\right) \times \frac{\vec{m}_p^{j+1} + \vec{m}_p^j}{2}\right] \tag{8}$$

where $\vec{m}_k^{(av)} = \dfrac{\vec{m}_k^{j+1} + \vec{m}_k^{j}}{2}$, $k = 1, 2, ... N$, $N$ - number of magnetic moments in a system, $\Delta\tau = \tau_{j+1} - \tau_j$, $j$ - counts the time instant.

Equation (8) can be written as

$$\left(\vec{m}_p^{j+1}\right)_i = \sum_l f_{il}(\vec{\Omega}_p) \cdot \left(\vec{m}_p^{j}\right)_l \tag{9}$$

Here $i, l = x, y, z$, $f_{il}(\vec{\Omega}_p)$ - are nonlinear functions of $\vec{\Omega}_p$. As parameters $\vec{\Omega}_p$ depend on $\vec{m}_k^{j+1}$ the solution of equations of motion was found by iterations.

In our simulations, we consider the system of identical cells (see Fig.1). Every cells contains $N = n^3$ spins located near the points of the cubic lattice:

$$\vec{r}_p = a(l_{p1}\vec{i} + l_{p2}\vec{j} + l_{p3}\vec{k}) + \delta\vec{r}_p.$$

Here $n$ is an odd number, $l_p$ - are integers, $-(n_{cell}-1)/2 \le l_{p1,2,3} \le (n_{cell}-1)/2$, $\delta x_p$, $\delta y_p$, $\delta z_p$ - are the random numbers, which do not exceed $0.05a$, $\vec{i}, \vec{j}, \vec{k}$ - are the unit vectors in the positive $x-$, $y-$, and $z-$directions. At $\tau = 0$ the $z-$components of $\vec{m}_p$ take random values between $-0.8$ and $-1$ with the random direction of the transversal components. In every cell, the central spin is the only FRS. In our simulations, we take the same value $\xi'_p = \beta$ for all FRS and the same value $\xi'_p = \alpha$ for all SRS.

The main (central) cell is surrounded by 440 identical auxiliary cells in order to eliminate the boundary effects, which influence the relaxation process. We computed the dynamics of the spins in the central cell taking into consideration their interaction with the spins of all the cells and assuming that the corresponding spins in all the cells have the same direction. The outcome of our computation is the function

$$M_z(\tau) = \sum_p m_{pz}(\tau) \bigg/ \sum_p |m_{pz}(0)| \tag{10}$$

We have performed computations for $n = 5, 7, 11$. The corresponding ratio $n_s / n_f = N - 1 = 124, 342, 1330$.

### 3. Results of numerical simulations

First, we have studied the relaxation process in the absence of magnetic field gradient ($G' = 0$). We define the effective dimensionless relaxation time $\tau_1$ using the relation

$$\tau_1 = \frac{1}{2}\int_0^\infty [1 - M_z(\tau)]\, d\tau, \tag{11}$$

which is derived from the exponential decay $M_z(\tau) = 1 - 2\exp(-\tau/\tau_1)$. As $\xi_0 \omega_0 = 1 s^{-1}$, the numerical value of $\tau_1$ is equal to the value $T_1$ in seconds. For the spin system with no FRS the relaxation time $\tau_1 = \tau_{SL} \approx 2.2/\alpha$. In experiment [7] $\tau_{SL} = 25s$, and $\alpha = 0.088$. We have simulated the relaxation process for various values $\chi > 10^4$, but all our figures below correspond to the experimental value $\chi = 3.2 \times 10^5$.

Our computations show that the overall relaxation time $\tau_1$ can be described by the relation

$$\tau_1 = \frac{2.2}{\alpha + \beta/N}. \tag{12}$$

Putting $N \approx n_s / n_F$ and transferring to the dimensional relaxation rate, we obtain from (12)

$$T_1^{-1} = T_{SL}^{-1} + T_0^{-1}. \tag{13}$$

The first term in this expression describes the spin-lattice relaxation rate for SRS, and the second term $T_0^{-1} = (n_F / n_s) T_{FL}^{-1}$, first derived in [5] (see Introduction), describes the effect of the spin diffusion. From the experimental data [7] $T_1^{-1} = 0.16 s^{-1}$, $T_{SL}^{-1} = 0.04 s^{-1}$ we can estimate the value of $T_0^{-1}$: $T_0^{-1} = 0.12 s^{-1}$. The corresponding value $\beta/N \approx 0.26$.

Fig. 2 demonstrate the dependence $M_z(\tau)$ for various values of the ratio $\beta/N$. Fig. 3 shows the dependence of the relaxation time $\tau_1$ on the ratio $\beta/N$ for $\alpha = 0$ and $\alpha = 1$. One can see the excellent agreement between the numerical data and formula (12).

Next, we transfer to the main objective of the paper: analysis of the influence of the magnetic field gradient on the relaxation process. Our simulations show that the suppression of the relaxation rate depends on the single parameter $K = G'/\chi$, which is the ratio of the characteristic Larmor frequency difference to the dipole-dipole interaction constant. The two other results of our simulations are the following:

1) The significant increase of the relaxation time $\tau_1$ appears in the region $4.4 \leq K \leq 44$, which corresponds to the values $10^3 T/m \leq G \leq 10^4 T/m$ in a good agreement with experiments [7].

2) For $K > 44$, the ratio $R(K) = \tau_1(K)/\tau_1(K=0)$ is approaching to the value $1 + \beta/(\alpha N)$. It means that the overall relaxation time $\tau_1(K)$ is approaching to the expected value $\tau_{SL}$.

As an example, Fig. 4 demonstrates dependence $M_z(\tau)$ for 5 values of $K$, and Fig. 5 shows the function $R(K) = \tau_1(K)/\tau_1(K=0)$.

Finally, we have studied the dynamics of the spatial distribution $m_z(\vec{r},t)$. We have found that in the absence of the magnetic field gradient ($K=0$) the relaxation process spreads randomly in all directions from FRS to SRS.

In the presence of the magnetic field gradient the spin diffusion process becomes anisotropic. Fig. 6 demonstrates the relaxation of a slice magnetic moment $M_{zi}$ for all the slices, $-(n-1)/2 \leq i \leq (n-1)/2$, in the central cell. One can see that the relaxation process first develops in the slice containing FRS ($i=0$), then it spreads to the slices $i<0$ below the central slice, then it spreads to the upper slices $i>0$. This phenomenon can be explained as following.

The spin frequency in the rotating frame is given by eq. (4). We have

$$\Omega_p \approx K\chi(z_p/a) + \Omega_{pd}, \quad \Omega_{pd} = \chi \sum_{k \neq p} A_{kp} m_{kz}, \tag{14}$$

where the first term, $\sim K$, is associated with the nonuniform magnetic field, and the second term, $\Omega_{pd}$, is associated with the dipole field produced by the other spins of the system.

Within the slice the value of $\Omega_{pd}$ changes slightly between $5.6\chi$ and $6.4\chi$. Thus, the value $\Omega_p$ is, approximately, $\Omega_0 = 6\chi$ in the slice $i=0$, $\Omega_1 = (6+K)\chi$ in the slice $i=+1$, $\Omega_{-1} = (6-K)\chi$ in the slice $i=-1$ and so on. It is clear that for $\Omega_1 - \Omega_0 = \Omega_0 - \Omega_{-1} \geq \Omega_0$ (or $K > 6$) the relaxation due to the spin diffusion develops first in the slice containing FRS ($i=0$). When the magnetic moment $M_{z0}$ in the central slice, $i=0$, changes its direction, $M_{z0} = -1 \rightarrow M_{z0} = +1$, the dipole contribution on the neighboring slices $i = \pm 1$ in our disturbed cubic lattice increases as

$$\Delta \Omega_i \approx 0.35\chi [1 + M_{z0}(\tau)], \quad i = \pm 1 \tag{15}$$

The dipole term $\Omega_{dp}$ within the central slice itself changes as

$$\Delta \Omega_0 \approx -9\chi [1 + M_{z0}(\tau)]. \tag{16}$$

Thus, after the spin relaxation in the slice $i=0$ the frequency difference between the slices decreases for the slices $i=0$ and $i=-1$, but increases for $i=0$ and $i=1$. As a result the spin relaxation process due to spin diffusion starts in the slice $i=-1$, then in the slice $i=-2$, and so on. The characteristic value $K = K_0$, which is necessary to suppress the spin diffusion between the slices, can be estimated as following. We assume that at $K = K_0$ the frequency difference between the slices $i=0$ and $i=-1$ (after the spin relaxation in the slice $i=0$) remains greater than $\Omega_0 = 6\chi$. Taking into consideration that the initial frequency difference between these slices is $K_0 \chi$, the change of the frequency is $-18\chi$ for $i=0$ and $0.7\chi$ for $i=-1$ we obtain $K_0 \approx 25$, which is in a good agreement with our numerical simulations.

## 4. Conclusion

We simulated the process of electron spin diffusion and relaxation in the presence of a nonuniform magnetic field using the quasiclassical equations of motion. The dipole-dipole interaction between the spins causes the energy transfer from SRS to FRS greatly increasing the relaxation rate. The Larmor frequency difference caused by the magnetic field gradient suppresses the spin diffusion and recovers the long relaxation time in the spin system.

The main results of our simulations are the following:

1. In the absence of the magnetic field gradient we proved the equation (13) for the rate of the spin relaxation $T_1^{-1}$ and computed the value of parameter $T_0 = (n_F / n_S) T_{FL}^{-1} \approx 0.12 s^{-1}$ for experiment [7].

2. We have shown that the suppression of spin diffusion in the presence of the magnetic field gradient depends on a single parameter, which is the ratio of the characteristic Larmor frequency difference to the dipole-dipole interaction constant. We have demonstrated a good agreement between the numerical simulations and experiment [7] on the spin diffusion suppression.

3. We have shown that the spin diffusion becomes highly anisotropic in the presence of the magnetic field gradient. The relaxation process first develops in the transversal slice containing FRS, then it spreads in the direction of the smaller magnetic field, and, finally, in the region of higher magnetic field.

## Acknowledgement

We thank Dan Rugar for useful discussions. This work was supported by the National Security Agency (NSA) and Advanced Research and Development Activity (ARDA) under Army Research Office (ARO) contract # 707003.

## Fig. captions

Fig.1 A pattern of the $x-y$ plane containing FRS for $n=7$. The values of $x$ and $y$ are given in units of $a$. $\bullet$ – FRS.

Fig.2. Spin relaxation $M_z(\tau)$ for the various values of the ratio $\beta/N$ and $\alpha=1$. Curves 1-7: $\beta/N = 0, 1, 2, 3, 4, 5, 6$.

Fig.3. The overall relaxation time $\tau_1$ as a function of ratio $\beta/N$: 1-$\alpha=1$, 2-$\alpha=0$. Dots are received from the numerical calculations, solid line corresponds to formula (12).

Fig.4. The relaxation $M_z(\tau)$ for various values of $K$: 1-5 - $K=0, 5, 10, 20, 50$. Curve 6 is the relaxation in the spin system with no FRS (and $K=0$). The values of other parameters: $\alpha=1$, $n=7$, $\beta=1372$ ($\beta/N=4$).

Fig.5. Dependence $R(K)$ on $K$. at $\alpha=1$, $n=7$. Curves 1, 2 - $\beta=1372$, 686 ($\beta/N=4, 2$).

Fig.6. Relaxation of the slice magnetic moment $M_{zi}(\tau)$ in the presence of the magnetic field gradient: $a$ - for the central slice, which contain FRS ($i=0$), and for the slices below the central slice; $b$ -for the slices above the central slice. The values of parameters: $n=7$, $\alpha=1$, $\beta/N=4$, $K=10$.

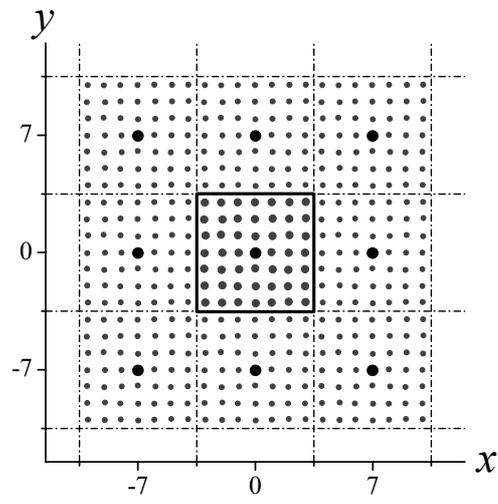

Fig. 1.

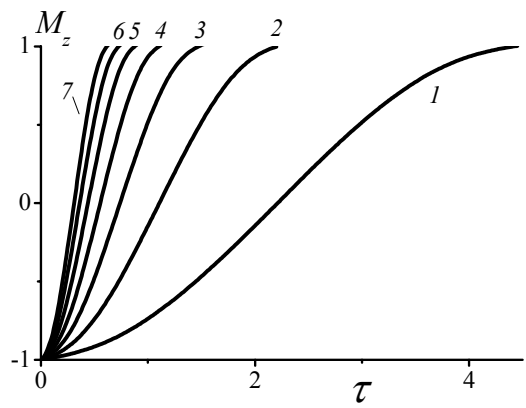

Fig. 2.

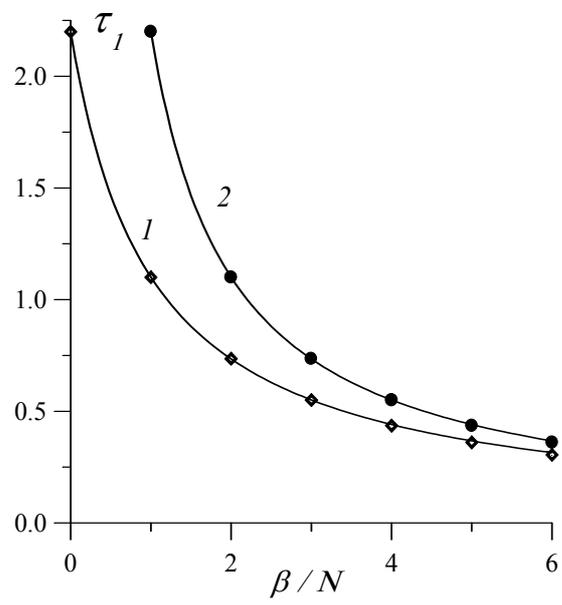

Fig. 3.

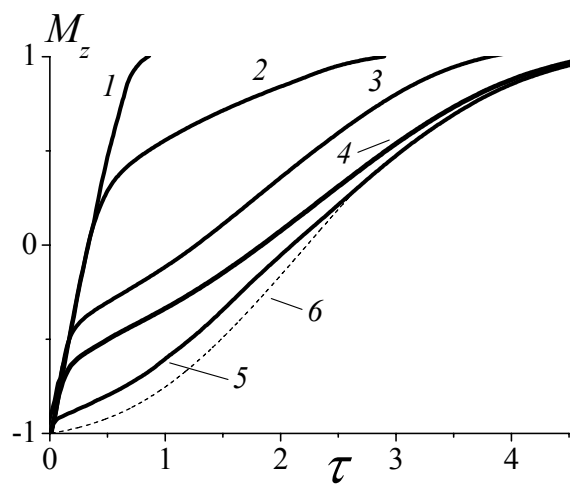

Fig. 4.

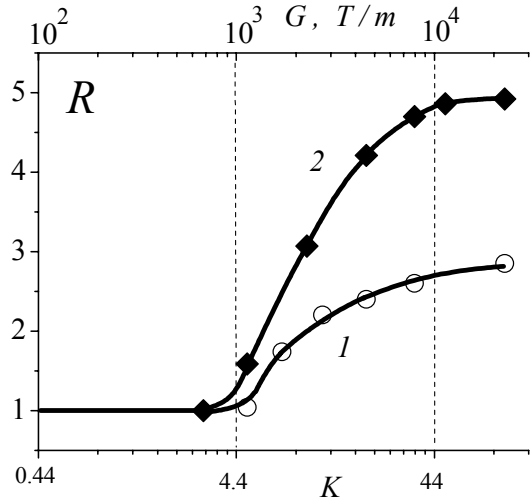

Fig. 5.

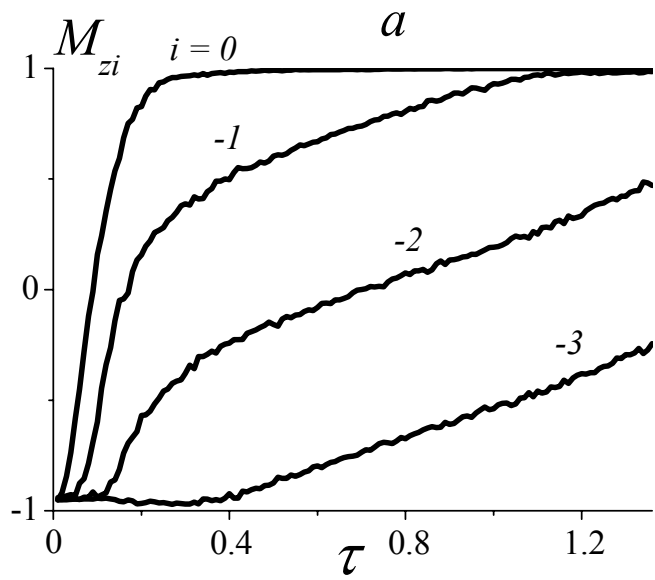

Fig. 6a.

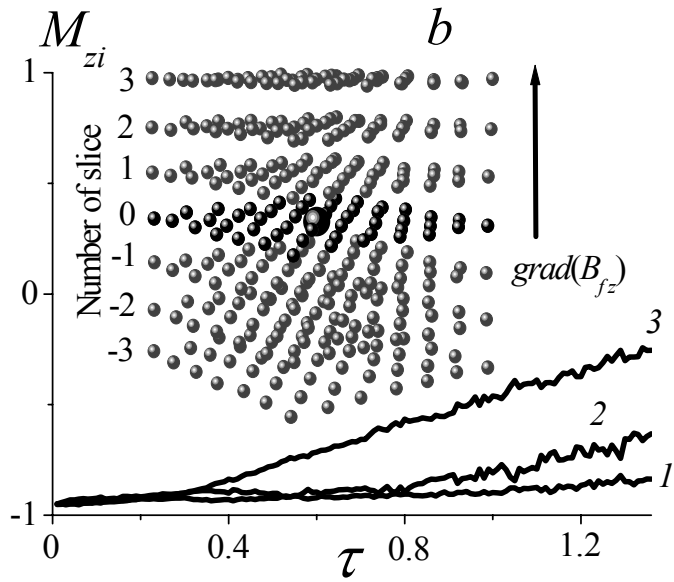

Fig. 6b.